\documentclass[10pt]{article}

\usepackage{times}
\usepackage{mathfont}
\def\log{\mathop{\rm log}}
\def\min{\mathop{\rm min}}

\DeclareSymbolFont{AMSb}{U}{msb}{m}{n}
\DeclareSymbolFontAlphabet{\Bbb}{AMSb}
\def\R{\ensuremath{\Bbb R}}

\usepackage{cite}

\usepackage{url}
\usepackage{graphics}

\newtheorem{lemma}{Lemma}
\newtheorem{corollary}{Corollary}
\newtheorem{theorem}{Theorem}

\newtheorem{conjecture}{Conjecture}
\newcommand{\qed}{~$\Box$\medbreak}
\newenvironment{proof}{\noindent{\bf Proof: }}{\qed}

\DeclareSymbolFont{lasy}{U}{lasy}{m}{n}
\let\Box\undefined
\DeclareMathSymbol\Box{0}{lasy}{"32}

\makeatletter
\long\def\@makecaption#1#2{
   \vskip 10pt 
   \setbox\@tempboxa\hbox{{\small #1. #2}}
   \ifdim \wd\@tempboxa >\hsize   
       {\small #1. #2}\par        
     \else                        
       \hbox to\hsize{\hfil\box\@tempboxa\hfil}  
   \fi}

\def\@begintheorem#1#2{\it\trivlist
				\item[\hskip \labelsep{\bf #1\ #2.\ }]}
\def\@opargbegintheorem#1#2#3{\it\trivlist
				\item[\hskip \labelsep{\bf #1\ #2\ {\rm(#3)}.}]}

\makeatother

\begin{document}

\title{Regression Depth and Center Points}

\urldef{\nahome}\url{http://www.cs.utexas.edu/users/amenta/}
\urldef{\namail}\url{amenta@cs.utexas.edu}
\urldef{\mbhome}\url{http://www.parc.xerox.com/csl/members/bern/}
\urldef{\mbmail}\url{bern@parc.xerox.com}
\urldef{\dehome}\url{http://www.ics.uci.edu/~eppstein/}
\urldef{\demail}\url{eppstein@ics.uci.edu}
\urldef{\sthome}\url{http://www-sal.cs.uiuc.edu/~steng/}
\urldef{\stmail}\url{steng@cs.uiuc.edu}

\def\aublock#1{\hbox to 1.5in{\hfil #1\hfil}}

\author{{\aublock{Nina Amenta\thanks{Univ. of Texas, Austin, Dept. of
Comp. Sci., \namail, \nahome}}\qquad
\aublock{Marshall Bern\thanks{Xerox Palo Alto Research Ctr., \mbmail,
\mbhome}}}\\ \\
{\aublock{David Eppstein\thanks{Univ. of California, Irvine, Dept.
of Inf. and Comp. Sci., \demail, \dehome}}\qquad
\aublock{Shang-Hua Teng\thanks{Univ. of Illinois, Urbana-Champaign,
Dept. of Comp. Sci., \stmail, \sthome}}}}

\date{ }
\maketitle   

\begin{abstract}
We show that, for any set of $n$ points in $d$ dimensions, there exists a
hyperplane with regression depth at least $\lceil n/(d+1)\rceil$,
as had been conjectured by Rousseeuw and Hubert.
Dually, for any
arrangement of $n$ hyperplanes in $d$ dimensions there exists a point
that cannot escape to infinity without crossing at least  $\lceil
n/(d+1)\rceil$ hyperplanes.
We also apply our approach to related questions on the
existence of partitions of the data into subsets such that a common
plane has nonzero regression depth in each subset, and to the
computational complexity of regression depth problems.
\end{abstract}

\section{Introduction}

Robust statistics \cite{HamRonRou-86, RouLer-87} 
has attracted much attention recently within the
computational geometry community
due to the natural geometric formulation of many of its
problems. In contrast to least-squares regression, in which
measurement error is assumed to be normally distributed, robust
estimators allow some of the data to be affected by completely arbitrary
errors.
Researchers in
this crossover area have developed algorithms for problems such as
center point construction \cite{ClaEppMil-IJCGA-96, JadMuk-DCG-94,
NaoSha-CCCG-90}, slope selection
\cite{BroCha-CGTA-98, ColSalSte-SJC-89, DilMouNet-IJCGA-92,
KatSha-IPL-93, Mat-IPL-91}, and the least median of squares regression
method \cite{EdeSou-JASA-90, MouNetRom-SODA-97} proposed by Rousseeuw
\cite{Rou-JASA-84}.

Recently, Rousseeuw and Hubert \cite{HubRou-JMA-98, RouHub-JASA-99,
RouHub-DCG-?} introduced {\em regression depth} as a
quality measure for robust linear regression: in statistical terminology,
the regression depth of a hyperplane $H$
is the smallest number of residuals that need to
change sign to make $H$ a nonfit. This definition has convenient
statistical properties such as invariance under affine
transformations; hyperplanes with high regression depth behave well
in general error models, including
skewed or heteroskedastic error distributions.

Geometrically, the regression depth of a hyperplane is
the minimum number of points intersected by the hyperplane as it
undergoes any continuous motion taking it from its initial position to
vertical.  In the dual setting of hyperplane arrangements, the
{\em undirected depth} of a point in an arrangement is the minimum
number of hyperplanes touched by or parallel to a ray originating at the
point.  Standard techniques of projective duality transform any statement
about regression depth to a mathematically equivalent statement about
undirected depth and vice versa.

Rousseeuw and Hubert \cite{RouHub-JASA-99, RouHub-DCG-?}
showed that for any $n$ and $d$ there exist sets of
$n$ points in $d$ dimensions such that no hyperplane has regression
depth larger than $\lceil n/(d+1)\rceil$.  For $d=2$, they found a simple
linear-time construction which achieves the optimal
$\lceil n/3\rceil$ bound. These facts, together with an analogy to {\em
center points} (points such that any halfspace containing them also
contains many data points), led to the following conjectures:

\medskip 

\begin{conjecture}[Rousseeuw and Hubert]\label{conj:depth}
For any $d$-dimensional set of $n$ points there exists a hyperplane having
regression depth
$\lceil n/(d+1)\rceil$.
\end{conjecture}

\begin{conjecture}[Rousseeuw and Hubert]\label{conj:partition}
For any point set there exists a partition into $\lceil
n/(d+1)\rceil$ subsets and a hyperplane that has
nonzero regression depth in each subset.
\end{conjecture}

Steiger and Wenger \cite{SteWen-CCCG-98} made some progress on
Conjectures \ref{conj:depth} and~\ref{conj:partition}: they show
that any point set can be
partitioned into $c_d n$ subsets, where $c_d$ is a constant
depending on the dimension~$d$, such that there exists a hyperplane
having nonzero regression depth in each subset.  Note that such a
hyperplane must have regression depth at least $c_d n$. Their value
$c_d$ is not stated explicitly, however it appears to be quite
small: roughly $1/(6^{d^2} (d+1))$.

Questions of computational efficiency of problems related to regression
depth have also been studied. Rousseeuw and Struyf \cite{RouStr-SC-98}
described algorithms for testing the regression depth of a given
hyperplane.  The same
paper also considers algorithms for testing the {\em location depth} of a
point (its quality as a center point).
One can find the hyperplane of greatest regression depth for a
given point set in time $O(n^d)$ by a breadth first
search of the dual hyperplane arrangement; standard
$\epsilon$-cutting methods \cite{MulSch-HDCG-97} can be used to develop a
linear-time approximation algorithm that finds a hyperplane with
regression depth within a factor
$(1-\epsilon)$ of the optimum in any fixed dimension.  For bivariate data,
van Kreveld, Mitchell, Rousseeuw, Sharir, Snoeyink, and Speckmann found
an algorithm for finding the optimum regression line in time $O(n\log^2
n)$ \cite{KreMitRou-SCG-99}, recently improved to $O(n\log n)$
by Langerman and Steiger \cite{LanSte-99}.

Our main result is to prove the truth of Conjecture
\ref{conj:depth}. We do this by finding a common generalization of
location depth and regression depth that formalizes the analogy between
these two concepts: the {\em crossing distance} between a point and a
plane is the smallest number of sites crossed by the plane in any
continuous motion from its initial location to a location incident to
the point. The location depth of a point is just its crossing
distance from the plane at infinity, and the regression depth of a plane
is just its crossing distance from the point at vertical infinity. We
then prove the conjecture by using Brouwer's fixed point
theorem to find a projective transformation that maps the point at
vertical infinity to a center point of the transformed sites; the inverse
transformation maps the plane at infinity to a deep plane.

We also improve the partial result of Steiger and Wenger on Conjecture
\ref{conj:partition}: we show that one can always partition a data set
into
$\lceil n/d(d+1)\rceil$ subsets with a hyperplane having nonzero
regression depth in each subset.
We further improve this to $\lfloor (n+1)/6\rfloor$ for $d=3$.
Our technique of projective
transformation also sheds some light on issues of computational
complexity: the two problems of testing regression depth and location
depth considered by Rousseeuw and Struyf
are in fact computationally
equivalent. Known NP-hardness results for center points then lead to the
observation that testing regression depth is NP-hard for data sets
of unbounded dimension.

\section{Overview of the Proof}

Before we begin the detailed proof, we describe our proof
strategy and outline some of the points of difficulty.

As discussed above, it is sufficient to find a projective
transformation such that the image of the point at vertical
infinity is a center point of the transformed set.
Equivalently, the point at vertical infinity should have large crossing
number with the plane at infinity of the transformed set, so the inverse
image of this plane has high regression depth.

To find such a transformation, we
view our space $\R^d$ as being embedded in $\R^{d+1}$,
tangent to a $d$-sphere, use central projection to lift the points
in $\R^d$ to pairs of points on the $d$-sphere, and use central
projection again to flatten them onto a copy of $\R^d$ tangent at a
different point $p$ of the $d$-sphere.
In this way, we get a different transformation for each point $p$ of the
sphere.  For each such transformation we consider a point $f(p)$ on
the sphere, found by computing a center point of the transformed
point set and lifting it back to the sphere again.
Note that $f(p)$ will automatically be in the same hemisphere as~$p$.

By the Brouwer fixed point theorem, any continuous function on the
sphere that maps points to the same hemisphere must be surjective
(Corollary~\ref{surj}).  If $f$ is surjective,
there exists a $p$ for which $f(p)$ is the lifted image of the point
at vertical infinity, giving us the transformation we want.

However, there are some technical difficulties.  As sketched above,
$f(p)$ is not continuous, for two reasons: first, there may be a large
set of center points, and it is difficult to pick a single one in a
continuous way. Second, and more importantly, as we move $p$
continuously on the sphere, the set of center points changes drastically
at those times when $p$ makes an angle of $\pi/2$ with a member of our
point set, so that the transformed image of the point moves out to
infinity in one direction and comes back in another.

To make the set of center points change more continuously, we approximate
the lifted point set on the sphere by a smooth measure. It is not
hard to generalize the concept of location depth to measures,
and to extend the proof of the existence of center points to this
setting (Lemma~\ref{ctrpt}), but there still may not exist a unique center
point. To chose a single continuously varying point
$f(p)$, we use the centroid of the set of points with location depth
$\ge\lceil n/(d+1)\rceil-\epsilon$.  Proving that this defines a
continuous function involves defining an appropriate metric on a space of
measures (Lemma~\ref{measure-metric}), representing $f(p)$ as a
composition of functions to and from this space of measures, and using
the fact that the set of points used to define $f(p)$ is convex with
nonempty interior (Lemma~\ref{epsilon-bounded}) together with smoothness
assumptions on the measure to show that the terms in this composition are
each continuous (Lemmas \ref{rotcont}, \ref{flatcont},
and~\ref{epscont}).

If we now apply the same Brouwer fixed point argument, we get a
transformation that takes the point at vertical infinity to a point with
location depth $\lceil n/(d+1)\rceil-\epsilon$.
This gives us a hyperplane $H$ with high, but not quite high enough,
regression depth in the measure approximating our point set.
To finish the argument, and prove the existence of a hyperplane with
high regression depth, we show that there exists an $\epsilon$,
and a measure approximating the point set and having the required
smoothness properties, such that we can find a hyperplane near $H$
with the stated bound on regression depth for the original point set
(Lemmas \ref{lem:sharp} and~\ref{lem:unsharp}).

\section{Geometric Preliminaries}

\subsection{Projective Geometry}

Although Rousseeuw and Hubert's conjectures are defined purely in terms
of Euclidean geometry, our proof fits most naturally in the context of
projective geometry.  We briefly review this geometry here, since
standard textbooks (e.g. \cite{Cox-87}) concentrate primarily on the
planar version, and we need higher dimensions.

Perhaps the simplest way to view $d$-dimensional projective space is as
a renaming of Euclidean objects one dimension higher.
Call a {\em projective point} a line through the origin of
$(d+1)$-dimensional Euclidean space, and a {\em projective hyperplane} a
hyperplane containing the origin of the same $(d+1)$-dimensional space. 
Then these projective points and hyperplanes satisfy properties
resembling those of $d$-dimensional Euclidean points and hyperplanes.
Indeed, one can embed Euclidean space into this projective space, in the
following way: embed a copy of $d$-dimensional Euclidean space as a
hyperplane in $(d+1)$-dimensional space, avoiding the origin
(so this hyperplane is not a projective hyperplane).
Then through any point of the $d$-dimensional space, one can draw a
unique line through it and the origin; that is, the Euclidean point
corresponds to a unique projective point.  Similarly, each hyperplane in
the $d$-dimensional space corresponds to a unique projective hyperplane.
However, there is one projective
hyperplane, and there are many projective points, that do not come from
Euclidean points and hyperplanes in this way; namely the
$(d+1)$-dimensional hyperplane through the origin parallel to the
$d$-dimensional space, and all $(d+1)$-dimensional lines contained in
that hyperplane. We call these projective objects {\em points at infinity}
and {\em the hyperplane at infinity}.  In particular, all vertical
Euclidean hyperplanes, when extended to the projective space, meet in a
single projective point, which we call the {\em point at vertical
infinity} ($\hat\infty$ for short).

A {\em projective transformation} is a map from one projective space to
another of the same dimension that takes points to points, hyperplanes to
hyperplanes, and preserves point-hyperplane incidences.  These include
(extensions of) the usual Euclidean affine transformations, but also some
other transformations in which infinite points are mapped to finite
points or vice versa.

\subsection{Central Projection}

Central projection is a correspondence from hyperplanes to spheres
closely related to the extension described above from Euclidean to
projective spaces.

Suppose we are given a $d$-dimensional hyperplane $E$ in
$(d+1)$-dimensional space, tangent to a $d$-sphere $S$.
Then given any set $X$ of $n$ point sites in $E$, we can lift this set to
a set $\breve X$ of $2n$ point sites on $S$, as follows:
draw a line through each site and the center of $S$;
this line intersects $S$ in two points; place a site at both points.
Conversely, given any function $f:S\mapsto \R$,
we can ``flatten'' it to a function $\bar f:E\mapsto \R$, as follows:
for each point $x$ in $E$, draw a line through $x$ and the center of
$S$; this line intersects $S$ in two points $y$ and $z$, one of which
(say $y$) is in the open hemisphere of $S$ centered on the point of
tangency; let $\bar f(x)=f(y)$.
In either case we define the {\em pole} of the projection to be the
common point of tangency between the hyperplane and the sphere.

The effect of lifting a hyperplane to a sphere and
then flattening the sphere to a different hyperplane can be viewed as a
projective transformation: if one places the origin at the sphere
center, the operations of drawing a line through a point, as used in both
lifting and flattening, are exactly the way we embedded Euclidean space
in projective space as described earlier.  The two different hyperplanes
simply form different Euclidean views of the same projective
space.

If one is given a Euclidean space (without a tangent sphere) the act of
lifting to a sphere requires an arbitrary choice: where to put the
tangent sphere.  Similarly if one is given a sphere (without a tangent
hyperplane) the act of flattening to a hyperplane requires a choice of
where to put the pole, and is completely determined once that choice is
made. In our proof, we will find a projective transformation from one
space to another by choosing arbitrarily a tangent sphere to our initial
space, and then considering all possible pole locations on that sphere.

\subsection{Measure Theory}

A {\em measure} on a topological space $X$ is just a function $m$ from
a family of subsets of $X$ (which must be closed under the complement
and countable union operations, and include all the open and closed
subsets) to nonnegative real numbers, satisfying the property of {\em
countable additivity}: if a set
$S$ is a disjoint union of countably many measurable subsets, then
the measures of those subsets must form a convergent series summing
to~$m(S)$.  We restrict our attention to measures for which the
measurable sets are just the {\em Borel sets}: sets that can
be formed from open sets by a sequence of complement and countable union
operations.

The usual Euclidean volume (Lebesgue measure) in $\R^d$ is not quite a
measure under our definition, because we want even the whole space to
have finite measure, but it is a measure on any restriction of $\R^d$ to
a bounded subset, or on the surface of a sphere. One can also define a
{\em discrete measure} from a set of point sites, in which the measure of
a set $S$ is simply the number of sites it contains.

Any measure $m$ on a sphere can be flattened to a measure $\bar m$ on
Euclidean space: given a set $S$ in Euclidean space, let $\breve S$ be the
copy of $S$ lifted by central projection to a subset of the open
hemisphere centered on the pole of projection, and let
$\bar m(S)=m(\breve S)$.

We define a {\em smooth measure} $m$ on the $d$-sphere to be one
for which there is a bound $b$
such that, for any set $S$, $m(S)$ is at most $b$ times the
Lebesgue measure. We define a smooth measure on $\R^d$ to be one
formed by flattening a smooth measure on the sphere.
(This is stronger than simply requiring a bounded ratio between the
measure and Lebesgue measure in $\R^d$.) 
Since any Lebesgue measurable set is the difference of a countable
intersection of open sets with a measure-zero set \cite[Theorem
3.15]{Oxt-80}, a smooth measure is completely determined by its behavior
on open sets. We define a measure to be {\em nowhere zero} if all open
sets have nonzero measure; note that we do not require the measure of
the open sets to be bounded below by a constant times their Lebesgue
measure.

For any smooth measures $m_1$ and $m_2$ on the sphere or $\R^d$
define the {\em distance} between $m_1$ and $m_2$
to be the supremum of $|m_1(S)-m_2(S)|$ where $S$ ranges over all {\em
convex} subsets of $X$ (convex subsets of the sphere are defined to be
sets that can be flattened to a convex subset of $\R^d$).

\begin{lemma}\label{measure-metric}
The distance defined above is a metric on the space of smooth measures.
\end{lemma}

\begin{proof}
The distance is clearly symmetric.
Any open set can be decomposed into a union of countably many convex
sets, and we can use inclusion-exclusion to express its measure as a
series each term of which is the measure of a convex set; therefore
any two distinct smooth measures have nonzero distance.
The triangle inequality is satisfied separately by the values
$|m_1(S)-m_2(S)|$ for each $S$, so it is satisfied by the overall
distance as well.
\end{proof}

\begin{lemma}\label{rotcont}
Let $m$ be a smooth measure on a sphere, and let $R$ be the
group of rotations of the sphere.  Define the measure
$m_\rho(S)=m(\rho(S))$ for any $\rho\in R$.  Then the map from $\rho$ to
$m_\rho$ is a continuous function from $R$ to the space of smooth
measures.
\end{lemma}

\begin{proof}
We need to show that for any $\rho$ and $\epsilon$ we can find a
$\delta$ such that all rotations within $\delta$ of $\rho$ are
mapped to a measure within $\epsilon$ of $m_\rho$.
By symmetry of the space of rotations, we can assume $\rho$ is the
identity.

For any set $S$ and rotation $\theta$,
$|m(S)-m(\theta(S))|\le m(S\oplus\theta(S))=O(b|\theta|L)$,
where $b$ is the bound on $m$ in terms of Lebesgue measure assumed in
the definition of smoothness and $L$ is the Lebesgue measure of the
boundary of $S$.  For any convex set, $L$ is bounded independently of
$S$ by the measure of the equator of the sphere, so if we choose
$\delta=O(1/b)$, any rotation amount smaller than $\delta$
will have $|m(S)-m_\rho(S)|<\epsilon$
as desired.
\end{proof}

\begin{lemma}\label{flatcont}
Flattening a sphere to a hyperplane (with a fixed pole of projection)
induces a continuous map from the space of smooth measures on the sphere
to the space of smooth measures on $\R^d$.
\end{lemma}

\begin{proof}
Flattening can only decrease the distance between two measures, since
the flattened distance is of the same form (a supremum of values
$|m_1(S)-m_2(S)|$) but with fewer choices for $S$
(only those convex subsets of the sphere that are contained in a
particular open hemisphere).
\end{proof}

\subsection{Smoothing and Sharpening}

In order to avoid complicated limit arguments, we will approximate
the discrete measure of a set of sites by a single smooth measure,
carefully chosen so that we can translate halfspaces in one measure to
halfspaces in the other in a way that preserves the measure of
the cuts appropriately.
As a notational convention, we will use accented letters like $H'$ to
refer to objects related to the discrete measure, and unaccented
letters like $H$ to refer to the corresponding objects for the smooth
measure.

Any pair of hyperplanes in a projective space divides the space
into two subsets; we define a {\em double wedge} to be the closure of
any such subset.  In particular a Euclidean halfspace is a
special case of a double wedge in which one of the hyperplanes is the
hyperplane at infinity. 
Given a set of sites, we say that two hyperplanes are {\em combinatorially
equivalent} if they bound a double wedge that has no sites in its
interior.  Note that this is not really an equivalence relation because
of the possibility of sites on the boundary of the wedge.

The proofs of some lemmas in this section rely on {\em projective
duality}: in any projective space, one can find a correspondence
between it and a {\em dual space} of the same dimension, in which each
point $p$ corresponds to a dual hyperplane $p^*$, and each hyperplane
$H$ corresponds to a dual point $H^*$, such that $p$ is incident to $H$
if and only if $p^*$ is incident to $H^*$.
Note that, under this correspondence, the set of hyperplanes
passing within distance
$\delta$ of point $p$ is transformed into a set of points within some
neighborhood of hyperplane $p^*$.
 
\begin{lemma}\label{lem:sharp}
For any finite set of sites in $\R^d$, there exists a $\delta$
such that any hyperplane $H$ can be replaced by a combinatorially
equivalent hyperplane $H'$, such that $H'$ is incident to all sites
within distance $\delta$ of $H$.
\end{lemma}

\begin{proof}
Let $\delta$ be smaller than half the height of any nondegenerate
simplex formed by $d+1$ points. For any $H$, let $S_0$ denote the set of
sites within distance $\delta$ of $H$; then $S_0$ must be coplanar.  Let
$H_0$ be any plane incident to all sites in $S_0$,
and continuously rotate $H$ towards $H_0$ around an axis where the two
hyperplanes intersect. (This motion is easier to understand in the dual:
it is just motion along a straight line segment from $H^*$ to $H_0^*$.)
Note that with such a motion, the distance from $H$ to any point of
$H_0$, and in particular to any of the sites in
$S_0$, is monotonically decreasing, so no site can leave set $S_0$.
However, $H$ may move to within distance $\delta$ of some site $x$
outside of $S_0$; if this happens, we stop moving towards $H_0$, form set
$S_1=S_0\cup\{x\}$, find a plane $H_1$ incident to all points in $S_1$,
continue rotating towards $H_1$, etc.
Since there are only finitely many sites, this process must eventually
terminate with a plane $H'$ incident to all sites crossed by the motion
of $H$; therefore there are no sites interior to the
double wedge defined by $H$ and $H'$.
\end{proof}

\begin{lemma}\label{lem:unsharp}
For any finite set of sites in $\R^d$, there exists a $\delta$ such that
any hyperplane
$H'$ can be replaced by a combinatorially equivalent hyperplane $H$,
such that $H$ is at distance at least $\delta$ from any site.
\end{lemma}

\begin{proof}
Form the hyperplane arrangement dual to the set of sites; choose
$\delta$ small enough that each cell of the arrangement has a point not
covered by any $\delta$-neighborhood of any hyperplane.
For any $H$, let ${H'}^*$ be an uncovered point in a cell containing
$H^*$, and let $H'$ be the hyperplane dual to ${H'}^*$.
\end{proof}

We will apply Lemma~\ref{lem:sharp} to the original sites,
and Lemma~\ref{lem:unsharp} to their vertical projections.

\subsection{Center Points}

If we are given a set of point sites in $\R^d$,
the {\em location depth} (also known as {\em Tukey depth}) of a point $x$
(which may not necessarily be itself a site)
is defined to be the minimum, over all projections $\pi:\R^d\mapsto\R$, of
the number of sites with $\pi(s)\le\pi(x)$.
Equivalently, it is the minimum number of sites contained in any
closed halfspace containing $x$. (The halfspace corresponding to
projection $\pi$ is $\{y:\pi(y)\le\pi(x)\}$.)

More generally, if $m$ is a measure on $\R^d$,
we define the
the {\em location depth} of $x$ to be the minimum measure of any
halfspace containing $x$.
Note that for any $D$ the set of points with location depth at least
$D$ is an intersection of closed halfspaces, and is therefore closed and
convex.

A {\em Tukey median} is a point with
maximum location depth.  A {\em center point} is a point with location
depth at least $m(\R^d)/(d+1)$.
As is well known \cite{Bir-PCPS-59, DanGruKle-SPM-63, Rad-JMLS-46} a
center point exists for any discrete measure; equivalently, any Tukey
median is a center point. We extend this to arbitrary measures using the
main idea from one proof of the discrete case: applying
Helly's theorem to a family of high-measure sets.

\begin{lemma}\label{ctrpt}
For any measure $m$ on $\R^d$, there exists a
point with location depth at least $m(\R^d)/(d+1)$.
\end{lemma}

\begin{proof}
For any positive integer $i$, let $\epsilon=1/i$, let $X$ be a compact
convex set with measure at least $(1-\epsilon)m(\R^d)$
(such a set always exists since $\R^d$ is a countable union of
nested convex bounded sets) and consider the family $F$ of compact convex
subsets of $X$ with measure at least $(d/(d+1) + \epsilon)m(\R^d)$.
The measure of the complement in $X$ of any set in $F$ is at most
$m(X)/(d+1) - \epsilon m(R^d)$,
so the intersection of any $(d+1)$-tuple of
sets in $F$ must be nonempty.  By Helly's theorem \cite{Hel-JDMV-23,
DanGruKle-SPM-63}
there is a point $x_i$ contained in all members of $F$.

If any open halfspace disjoint from $x_i$ has measure larger than
$(d/(d+1) + 2\epsilon)m(\R^d)$, some closed halfspace contained in it also
has measure larger than
$(d/(d+1) + 2\epsilon)m(\R^d)$, and would intersect $X$ in a compact
convex set of measure larger than $(d/(d+1) + \epsilon)m(\R^d)$,
contradicting the assumption that $x_i$ is in the intersection of all
such sets. Therefore, $x_i$ has location depth at least
$(1/(d+1) - 2\epsilon)m(\R^d)$.

Since we can make $\epsilon$ as small as we wish (and since all points
with location depth at least
$\epsilon m(\R^d)$ must be contained in the compact set
$X$), we can find a cluster point of the points $x_i$,
and this cluster point must have location depth at least
$m(\R^d)/(d+1)$.
\end{proof}

Define an {\em $\epsilon$-center point} to be a point with location
depth at least $m(\R^d)/(d+1) - \epsilon$.

\begin{lemma}\label{epsilon-bounded}
For any smooth measure $m$ in $\R^d$, and any sufficiently small
$\epsilon$, the set of
$\epsilon$-center points is compact and has nonempty interior.
\end{lemma}

\begin{proof}
Let $m$ be formed by flattening a smooth measure $\breve m$ on the sphere,
and let $c$ be a center point of $m$. For any
$\epsilon$ there exists a
$\delta$ for which any infinite strip of width $\delta$ containing
$c$ has measure at most $\epsilon$ (since the lift of such a strip is a
narrow wedge of the sphere).
Therefore, the points in an open ball of radius $\delta$ around $c$ are
all
$\epsilon$-center points.

Let $\epsilon$ be small enough that the location depth of an
$\epsilon$-center point is bounded away from zero. The set of
$\epsilon$-center points is clearly closed by its definition. To show
that the set is bounded, note that for any
$\delta$ one can find a neighborhood of the equator of the sphere with
measure at most
$\epsilon$; for any point $x$ in this neighborhood, one can find a
halfspace in $\R^d$ containing the point that is a subset of the
flattening of this neighborhood, and that therefore has too small a
measure for $x$ to be an $\epsilon$-center point.  The complement of
this flattened neighborhood is a bounded region of $\R^d$.
\end{proof}

Define the {\em $\epsilon$-trimmed mean} of a measure $m$ to be the centroid
of its set of $\epsilon$-center points.

\begin{lemma}\label{epscont}
For any sufficiently small $\epsilon>0$, the map from measures to
$\epsilon$-trimmed means defines a continuous function from nowhere zero smooth
measures to $\R^d$.
\end{lemma}

\begin{proof}
Let $m$ be a smooth nowhere zero measure, and $K$ its set of
$\epsilon$-center points.  Then $K$ is an intersection of closed halfspaces,
so any point $x$ outside $K$ is contained in an open halfspace $H$ tangent
to $K$ and having measure $m(\R^d)/(d+1)-\epsilon$.
Let $S$ be the infinite slab bounded on one side by the boundary of
$H$, and on the other side by a hyperplane through $x$.
The halfspace on the other side of this slab from $K$
has measure $m(\R^d)/(d+1)-\epsilon-m(S)$,
and $x$ can only become an $\epsilon$-center point of a measure with
distance at least $m(S)$ from $m$.  In other words, for any $x$
outside $K$ there is a $\delta=m(S)$ such that measures within distance
$\delta$ of $m$ do not have $x$ in their set of $\epsilon$-center points.

For any $y$ interior to $K$, let $S_i$ (for $i=1\ldots2^d$) be the
intersections with $K$ of a system of orthants centered at $y$.
Then any halfspace containing $y$ can be decomposed into a slab
containing one of the $S_i$ and a smaller halfspace containing a
boundary point of $K$; therefore by a similar argument to the one above,
there is a $\delta=\min\{ m(S_i) \}$ such that all measures within
distance
$\delta$ of $m$ have $y$ in their set of $\epsilon$-center points.

Thus an arbitrarily small change to the measure can only change the set
of $\epsilon$-center points in an arbitrarily small region near the
boundary of $K$, which can only make the centroid of $K$ change by an
arbitrarily small amount.
\end{proof}

\subsection{Brouwer's Theorem and Functions on Spheres}

The following well-known fact about functions on spheres is a simple
consequence of the Brouwer fixed point theorem, that any continuous
function from a closed topological disk to itself has a fixed point
\cite{Bro-MA-10,Bro-MA-12}.

\begin{lemma}
Let $f$ be a continuous non-surjective function from a $d$-sphere $S$ to
itself. Then $f$ has a fixed point.
\end{lemma}

\begin{proof}
Since $f$ is non-surjective, there is a point $x$ not covered by~$f$. 
Since $f$ is continuous, it avoids an open neighborhood $N$ of
$x$.  Then the restriction of $f$ to $S\setminus N$ is a continuous map
from a closed disk to itself, and hence by the Brouwer fixed point
theorem has a fixed point.
\end{proof}

\begin{corollary}\label{surj}
Let $f$ be a continuous function from a $d$-sphere to itself such that
for all $x$, $f(x)\neq -x$.  Then $f$ is surjective.
\end{corollary}

\begin{proof}
Apply the lemma to $-f$.
\end{proof}

\section{The Proof}

If $m$ is a measure on a projective space,
we define the {\em crossing distance}
$\chi_m(x,H)$ between a point $x$ and a hyperplane $H$
to be the minimum measure of any double wedge where one boundary
hyperplane is $H$ and the other contains $x$.
Intuitively, if $m$ is a discrete measure coming from a set of point
sites, $\chi$ measures the number of points that must be crossed by $H$
in any continuous motion of hyperplanes that moves $H$ until it
touches~$x$.

Then, the location depth of a point $x$ is simply $\chi_m(x,\infty)$
where $\infty$ denotes the hyperplane at infinity.
Conjecture~\ref{conj:depth} can be rephrased as asking for a hyperplane
$H$ such that $\chi_m(\hat\infty,H)$ is large.
Therefore, location depth and regression depth are both special cases of
crossing distance.

Since crossing distance is defined purely projectively, it is preserved
by any projective transformation.  Thus if we find a hyperplane with
high regression depth, performing a projective transformation that takes
it to the hyperplane at infinity will also take the point at vertical
infinity to a center point.  Conversely, if we can find a projective
transformation that takes the point at vertical infinity to a center
point, the preimage of the hyperplane at infinity must have high
regression depth.

Our proof of Conjecture~\ref{conj:depth}, below, finds such a
transformation as a composition of two central projections.
The idea of the proof is very simple: lift the sites to a sphere, flatten
the sphere at a pole, compute the center point of the flattened
points, and use Corollary~\ref{surj} to show that this map from poles to
center points covers $\hat\infty$. All of the technical complication in
the proof arises from our need to force ``the center point'' to be unique
and the map to be continuous, which we do by approximating the points with
smooth measures and using
$\epsilon$-trimmed means of these measures.

\begin{theorem}\label{thm:depth}
For any 
$n$ points in $\R^d$
there exists a hyperplane
having regression depth
$\lceil n/(d+1)\rceil$.
\end{theorem}

\begin{proof}
Use central projection with an arbitrary fixed choice of tangent
sphere to lift the sites to a set of $2n$ points on a sphere.
The extension of this lifting map to the projective space lifts
the point at vertical infinity to two points on the sphere; choose one of
these two arbitrarily and call it $\hat\infty$.

Let $\delta$ be small enough that we can apply Lemma~\ref{lem:sharp} to
the sites and Lemma~\ref{lem:unsharp} to the vertical projection of the
sites.
Let $\epsilon=1/3(d+1)$.  Choose a smooth nowhere zero measure
$m$ such that the measure of any hemisphere is $n$, the
$\delta$-radius ball around any site has measure at most one,
and the total measure of the set of points farther than $\delta$ from
any site is at most $\epsilon$.

Define the function $c(x)$ from the sphere to itself as
follows: let $\bar m(x)$ be the measure formed by flattening $m$
at pole $x$, let $\bar c(x)$
be the $\epsilon$-trimmed mean of $\bar m(x)$, and use central projection to
lift
$\bar c(x)$ to a point $c(x)$ in the open hemisphere centered on $x$.  By
Lemmas
\ref{rotcont}, \ref{flatcont}, and~$\ref{epscont}$, $c$ is continuous,
and clearly $c$ has no point for which $c(x)=-x$.
Then by Corollary~\ref{surj}, $c$ is surjective, so we can find a
point $p=c^{-1}(\hat\infty)$ such that flattening the sphere tangent
to $p$ maps $\hat\infty$ to the $\epsilon$-trimmed mean of $\bar m(p)$.

Let $H$ denote the hyperplane at infinity with respect to $p$.
Use Lemma~\ref{lem:sharp} to sharpen $H$ to a hyperplane $H'$ incident
to all sites within distance $\delta$ of $H$.
We wish to show that $H'$ has the stated
regression depth; that is, any double wedge bounded by $H'$ and a vertical
hyperplane $V'$ must contain at least $\lceil n/(d+1)\rceil$ sites.
Thus let $V'$ be an arbitrary vertical hyperplane, and let $W'$ be a
double wedge determined by
$H'$ and $V'$.  Use Lemma~\ref{lem:unsharp} to smooth $V'$ to a vertical
hyperplane $V$ that is not within distance $\delta$ of any site, and let
$W$ be the double wedge determined by $H$ and $V$. Then since $\hat\infty$
is an
$\epsilon$-trimmed mean for $\bar m(p)$,
$W$ has measure at least $n/(d+1)-\epsilon$, and
the measure of the
intersection of $W$ with the $\delta$-radius balls around the sites
must be at least
$n/(d+1)-2\epsilon$.  Therefore $W$ must contain or cross at least
$\lceil n/(d+1)-2\epsilon\rceil=\lceil n/(d+1)\rceil$ of the balls, and
$W'$ contains at least that many sites.
\end{proof}

\section{Analogues of Helly's Theorem}

\begin{figure}[t]
$$\includegraphics{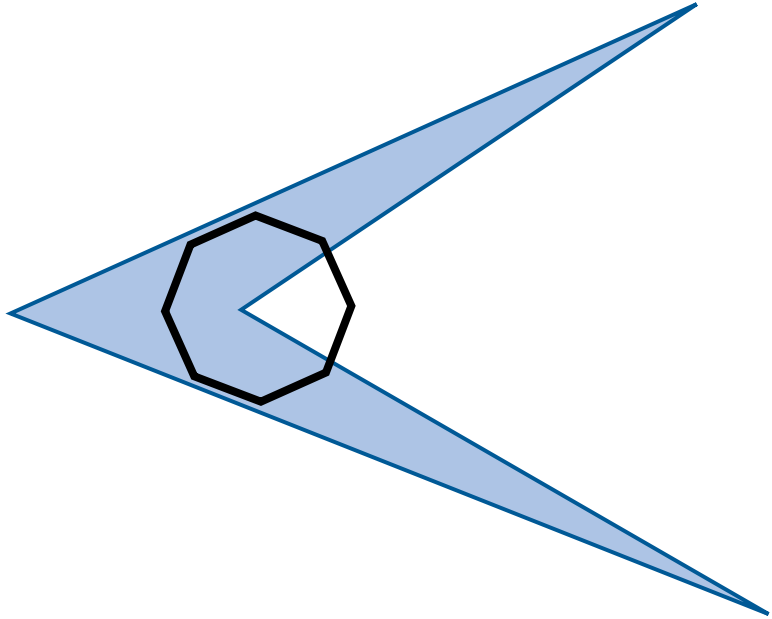}$$
\caption{One of a family of $n$ contractible hulls such that all sets of
$n-1$ hulls have a common intersection, but not all $n$ do.}
\label{anti-helly}
\end{figure}

Rousseeuw \cite{Rou-SCG-98} expressed the hope for an alternate proof
of Conjecture~1 analogous to that of Lemma~\ref{ctrpt}, based on
some formulation of Helly's theorem for {\em contractible hulls}
(sets of hyperplanes having nonzero regression depth for some point set).
The natural formulation is that, if every sufficiently large subset of a
family of contractible hulls has nonempty intersection, then the whole
family has nonempty intersection.  However, despite some formal
similarities between similarly defined shapes and convex polygons
\cite{EuGueTou-Algs-96}, there can be no such result, as we  now show.

We use the projective dual
formulation, in which the contractible hull of  an arrangement of lines
consists of those points not interior to an  infinite cell of the
arrangement.
Figure~\ref{anti-helly} shows how, for a regular $n$-gon, one can find a
set of four lines such that their contractible hull (the set of points
that cannot reach infinity, consisting of a nonconvex quadrilateral
together with the points on the lines themselves) contains all but one
$n$-gon vertex, does not contain the $n$-gon center, and has its two outer
lines perpendicular to the two
$n$-gon sides adjacent to the missed vertex.
Thus, the hull is completely disjoint from a wedge defined by two rays
emanating from the $n$-gon center, parallel to the hull's two outer
lines.   If we form $n$ of these hulls, one per $n$-gon vertex,
the union of the corresponding wedges is the entire plane; therefore the
intersection of the $n$ contractible hulls is empty.  However, any
subset of $n-1$ hulls do have a common
intersection, including at least the $n$-gon vertex missed by the one
hull not in the subset.

However, Rousseeuw (personal communication) noted that Theorem~\ref{thm:depth}
does imply some sort of special case of a Helly theorem: the contractible
hulls of all $(n d/(d+1))$-tuples of sites have a common intersection.
It remains unclear whether this can be formalized as a more general
Helly theorem for families of contractible hulls.

\section{Analogues of Tverberg's Theorem}

A {\em Tverberg partition} of a set of point sites is a partition of the
sites into subsets, the convex hulls of which all have a common
intersection.  (To extend this definition to the projective plane, we
define the convex hull of a point at infinity to be the whole plane.)
The {\em Tverberg depth} of a point $x$ is the maximum cardinality of any
Tverberg partition for which the common intersection contains $x$.
Note that the Tverberg depth is a lower bound on the location depth.
Tverberg's theorem \cite{Tve-JLMS-66, Tve-BAMS-81} is that there always
exists a point with Tverberg depth $\lceil n/(d+1)\rceil$ (a {\em
Tverberg point}); this result generalizes both the existence of center
points (since any Tverberg point must be a center point) and Radon's
theorem \cite{Rad-MA-21} that any $d+2$ points have a Tverberg partition
into two subsets.

Similarly, define a {\em contractible partition} of a set of point sites
to be a partition of the sites into subsets, the contractible hulls of
which all have a common intersection, and define the {\em contractible
partition number} of the set to be the maximum number of subsets in any
partition.  Conjecture~\ref{conj:partition} states that the contractible
partition number is always at least $\lceil n/(d+1)\rceil$.
Since a hyperplane $H$ is in the contractible hull of a set of
points if and only if a projective transformation taking $H$ to infinity
takes $\hat\infty$ to a point in the convex hull of the transformed set,
the contractible partition number is the maximum Tverberg depth
of the image of $\hat\infty$ under any projective transformation.
Thus the conjecture would be proven if we could find a projective
transformation taking $\hat\infty$ to a Tverberg point.

Unfortunately we have not been able to extend our previous proof to this
case.  We do not know of an appropriate generalization of Tverberg
points to continuous measures, and in any case Tverberg points are not
very well behaved: the set of Tverberg points need not be connected, if
it is connected it need not be simply connected, and in dimensions
higher than two its convex hull need not be the set of all centerpoints
\cite{Ten-PhD-91}.

However, we can at least show that the contractible partition number is
always at least $n/d(d+1)$, an improvement over the previous bound of
Steiger and Wenger \cite{SteWen-CCCG-98}:

\begin{lemma}
Let $c$ have location depth $D$ with respect to a set of $n$ sites.
Then $c$ has Tverberg depth at least $\lceil D/d\rceil$.
\end{lemma}

\begin{proof}
As long as $c$ is contained in the convex hull of the sites,
greedily choose some simplex with site vertices containing $c$ and remove
its sites from the set.  This process can continue until all sites in
some halfspace $H$ containing $c$ on its boundary have been removed.
Initially, $H$ has at least $D$ sites, and
each simplex can contain only $d$ points in $H$,
so at least $\lceil D/d\rceil $ simplices can be chosen before $H$ is
exhausted.
\end{proof}

\begin{theorem}
The contractible partition number is at least
$\lceil n/d(d+1)\rceil$.
\end{theorem}

\begin{proof}
Find a hyperplane $H$ of regression depth $\lceil n/(d+1)\rceil$ and
a projective transformation taking $H$ to the hyperplane at
infinity, and apply the lemma to the image of $\hat\infty$ under this
transformation.
\end{proof}

In two dimensions, the optimal bound $\lceil n/3\rceil$ was shown by
Rousseeuw and Hubert \cite{RouHub-JASA-99}, and a partition achieving
this bound can be found in linear time from their construction
\cite{HubRou-JMA-98}.

\section{Better Tverberg Partitions in Three Dimensions}

Our general result above implies that in three dimensions there always
exists a partition of the sites into $\lceil n/12\rceil$ subsets the
contractible hulls of which have a common intersection.
We now improve this bound
somewhat to $\lfloor(n+1)/6\rfloor$.

The idea behind our bound is to partition the
sites by a plane such that the two subsets, when projected onto a
horizontal plane, have equal centerpoints.  We will then be able to find
a Tverberg partition consisting of $\lfloor(n+1)/6\rfloor$ subsets, each
formed by a triangle above the partition plane and a triangle below the
partition plane, where the triangles come from an equivalence between
center points and Tverberg points in $\R^2$:

\begin{lemma}[Birch \cite{Bir-PCPS-59}]
Let point $x$ be a center point of a set of $3k$ sites in $\R^2$.
Then $x$ is also a Tverberg point for this set of sites.
\end{lemma}

The proof of Birch's result is simply to form $k$ triangles
by connecting every $k$th point in the sequence of sites sorted by their
angles around $x$.  We need the following strengthening of the lemma:

\begin{lemma}
Let point $x$ have location depth $k$ in a set of $n>3k$
sites in $\R^2$.  Then there is a subset of exactly $3k$ sites, such
that $x$ still has location depth $k$ in this subset.
\end{lemma}

\begin{proof}
Since $n\ge 3k+1$, and $k$ is an integer, $\lfloor(n-k-1)/2\rfloor\ge k$.
Let $H$ be a closed halfspace with $x$ on its boundary, containing
exactly $k$ sites. Sort the sites outside $H$ according to their angles
with $x$, and let $y$ be the median site in this sorted order.
Then the two closed wedges in the complement of $H$, bounded by line
$xy$, each contain at least $\lfloor(n-k-1)/2\rfloor\ge k$ sites,
not counting $y$.
If we remove $y$ from the set of sites, then the number of sites in any
halfspace not containing $y$ does not change, and any halfspace
containing $y$ contains one of these two wedges.
Therefore, the location depth of $x$ remains equal to $k$ and the result
follows by induction on $n$.
\end{proof}

\begin{corollary}\label{cor:ld=td}
Let point $x$ have location depth $k$ in $n$ point sites
in
$\R^2$. Then $x$ has Tverberg depth at least $\min\{k,\lfloor
n/3\rfloor\}$.
\end{corollary}

Given any oriented plane $P$ in $\R^3$, define 
$L(P)$ to be the closed halfspace to the left of $P$ (according to the
orientation of $P$) and $R(P)$ to be the closed halfspace to the right.
Let $\pi:\R^3\mapsto\R^2$ be a vertical projection from $\R^3$ to $\R^2$:
that is, $\pi(x,y,z)=(x,y)$.  Note that $\pi$ also acts as a continuous
function from smooth measures in $\R^3$ to smooth measures in $\R^2$,
according to the formula $\pi(m)(S)=m(\pi^{-1}(S))$.
If $S$ is any measurable set in $\R^3$, and $m$ is any measure on
$\R^3$, let $m\cap S$ denote the measure defined by the formula
$(m\cap S)(T)=m(S\cap T)$.

\begin{figure}[t]
$$\includegraphics{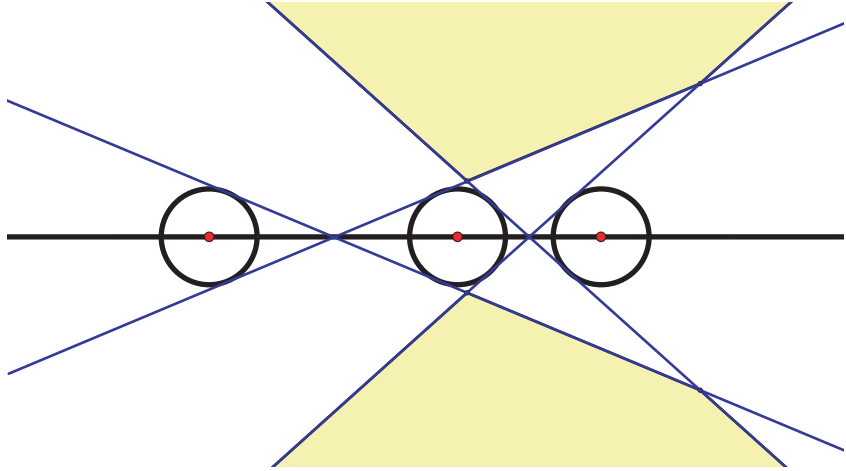}$$
\caption{The $\delta$-neighborhood of a line through $\ge 2$ sites.}
\label{L-nbhd}
\end{figure}

\begin{figure}[t]
$$\includegraphics{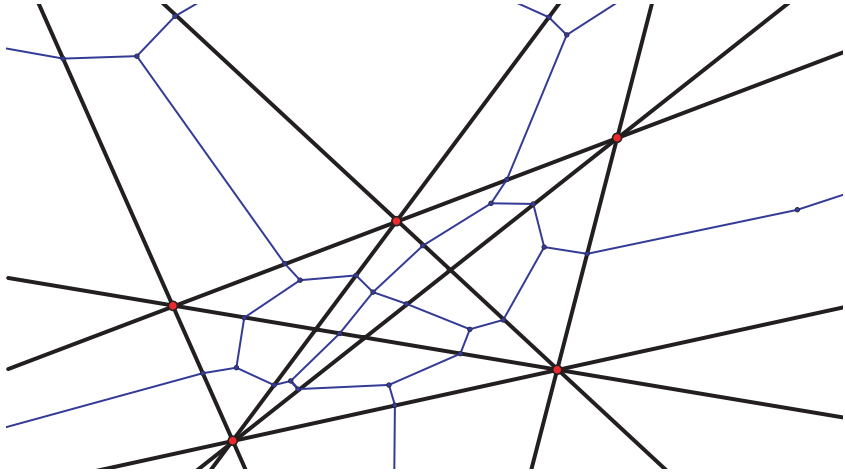}$$
\caption{Arrangement of lines determined by pairs of sites, and
subdivision of arrangement cells into quadrilaterals.}
\label{fig-c-unsharp}
\end{figure}

Given a set of point sites in $\R^2$, define points $c$ and $c'$ to be
{\em combinatorially equivalent} if there is no line determined by two
sites that has $c$ on one side and $c'$ on the other.
Define the {\em $\delta$-neighborhood} of a line $L$ through two or more
sites to be the set of lines determined by pairs of points within distance
$\delta$ of two distinct sites on $L$.
The lines of the $\delta$-neighborhood all lie within a region bounded by
two convex polygons, with sides formed by lines tangent to radius-$\delta$
circles around the sites on
$L$ (Figure~\ref{L-nbhd}). We say that a line $L$
determined by two sites
$p$ and
$q$ is {\em
$\delta$-near} $c$ if there is a line $L'$ through $c$ in the
$\delta$-neighborhood of $L$.  For any $c$ and $L$ not incident to $c$,
$L$ is not $\delta$-near~$c$ for all sufficiently small values
of~$\delta$.

\begin{lemma}\label{lem:c-unsharp}
For any finite set of sites in any bounded region of $\R^2$, there exists
a $\delta$ such that, for any point $c$ in the bounded region, we can find
a combinatorially equivalent point
$c'$, having the property that any line through two sites that is
$\delta$-near to $c$ passes through $c'$.
\end{lemma}

\begin{proof}
We first describe how to map $c$ to $c'$; we will then show that there
exists an appropriate $\delta$ for this map. Form the arrangement of all
lines through two or more sites, find a point $p_i$ interior to each cell
$C_i$ of the arrangement (other than infinite cells with only one
vertex), and divide
$C_i$ into small quadrilaterals by drawing line segments from
$p_i$ to the midpoints of the finite-length edges of $C_i$
(Figure~\ref{fig-c-unsharp}).
Within infinite cells of the arrangement,
we also add a ray from
$p_i$ to infinity, not parallel to either side of $C_i$. Our choice
of
$c'$ is determined by this subdivision: each point
$c$ interior to a quadrilateral is mapped to the unique arrangement
vertex $c'$ contained in that quadrilateral.  We can use an arbitrary
tie-breaking rule to assign points on the boundaries of quadrilaterals
to the arrangement vertex for any incident quadrilateral.

Now approximate the given bounded region by a square that contains it.
For any line $L$ determined by two sites, there exists a $\delta_L$
such that $L$ is not $\delta_L$-near any of the points $p_i$, arrangement
edge midpoints, arrangement vertices not incident to $L$,
or points where the square crosses one of the edges of the subdivision.
Each point $c$ in the bounded region that is not mapped to $L$ is
contained in the convex hull of some set of these points, all on the
same side of $L$. The complement of a $\delta$-neighborhood on one side
of $L$ is convex.  Therefore,
$L$ will not be $\delta_L$-near any point $c$ mapped to a
$c'$ not on $L$.  We simply choose $\delta$ to be the minimum of the
values~$\delta_L$.
\end{proof}

\begin{theorem}
The contractible partition number in $\R^3$ is at least
$\lfloor\lceil
n/2\rceil/3\rfloor=\lfloor(n+1)/6\rfloor$.
\end{theorem}

\begin{proof}
Let $\delta$ be small enough that we can apply Lemma~\ref{lem:sharp} to
the sites and Lemma~\ref{lem:c-unsharp} to the vertical projection of the
sites (with the bounded region of Lemma~\ref{lem:c-unsharp} being the
points within distance $\delta$ of the convex hull of the sites). Let
$\epsilon<1/18$, and find a smooth nowhere zero measure $m$ on
$\R^3$ such that the total measure is $n-2\epsilon$, the measure within
the radius-$\delta$ ball around any site is at most one, and the total
measure outside all such balls is at most $\epsilon$.
Let $\mu$ be a nowhere zero smooth measure on $\R^2$ with total
measure $\epsilon$.

For each unit vector $u$ in $\R^3$, let $P(u)$ denote the oriented plane
normal to $u$ for which $m(L(P))=m(R(P))$.  Note that $P(u)$ is unique
due to the assumption that $m$ is nowhere zero.
Let $f(u)$ denote the vector
difference between two points in $\R^2$: the
$\epsilon$-trimmed means of $\pi(m\cap L(P))+\mu$ and $\pi(m\cap
R(P))+\mu$. That is, if these two $\epsilon$-trimmed means have
Cartesian coordinates $(x_L,y_L)$ and $(x_R,y_R)$ then let
$f(u)$ be the vector $(x_L-x_R,y_L-y_R)$.
Then $f$ is a continuous antipodal function, so by the
Borsuk-Ulam Theorem \cite{Bor-FM-33} it has a zero $u$,
where the two $\epsilon$-trimmed means coincide at a common point~$c$.

Use Lemma~\ref{lem:sharp} to find a
plane $P'$ passing through all sites within distance $\delta$ of
$P(u)$; then $L(P')$ and $R(P')$ each contain at least
$\lceil n/2-\epsilon\rceil=\lceil n/2\rceil$ sites.
Use Lemma~\ref{lem:c-unsharp} to find a point $c'\in\R^2$
on any line $\delta$-near $c$.

Then $c'$ must have location depth at least $\lceil n/6\rceil$
with respect to each of the two
planar sets formed by vertically projecting $L(P')$ and $R(P')$.
For, let $h'$ be a closed halfplane with $c'$ on its boundary,
containing as few points as possible from $L(P')$ or $R(P')$;
let $k=\min\{|h'\cap L(P')|,|h'\cap R(P')|\}$.  Since $c'$ is not
$\delta$-near any line it is not incident to, we can rotate $h'$ if
necessary to a combinatorially equivalent halfplane such that the
boundary of $h'$ does not pass within distance $\delta$ of any
nonincident point. Next, translate the halfplane so that its boundary
moves from $c'$ towards $c$ without coming within distance $\delta$ of
any site outside $h'$.  If the halfplane gets stuck by becoming tangent
to a radius-$\delta$ circle around a site, rotate it towards $c$ while
keeping it tangent to that circle.  This rotation process can not become
stuck by hitting another such circle, because the two corresponding
sites would determine a line that either separates $c$ from $c'$ or is
$\delta$-near to $c$, neither of which can happen by
Lemma~\ref{lem:c-unsharp}.  So the result of this process must be a
halfplane
$h$, with boundary incident to $c$, that is at distance at least
$\delta$ from any site not in $h'$.  Therefore,
$h$ intersects the radius-$\delta$ circles around at most $k$ sites of
$L(P')$ or $R(P')$, so
$\min\{(\pi(m\cap L(P))+\mu)(h),(\pi(m\cap L(P))+\mu)(h)\}\le
k+2\epsilon$. But, since $c$ is an $\epsilon$-trimmed mean,
$\min\{(m\cap L(P)+\mu)(h),(m\cap L(P)+\mu)(h)\}\ge n/6-\epsilon$.
Therefore, $k\ge n/6-3\epsilon$, and, since $\epsilon<1/18$ and $k$ is
an integer, $k\ge \lceil n/6\rceil$.

By Corollary~\ref{cor:ld=td}, we can find a set $TL$ of
$\lfloor\lceil n/2\rceil/3\rfloor$
triangles having as vertices sites in $L(P')$, such that the projection
of each triangle contains $c'$, and a corresponding set $TR$
of $\lfloor\lceil n/2\rceil/3\rfloor$ triangles with vertices in~$R(P')$.

We now use these triangles to form contractible hulls containing $P'$.
Whenever some triangle has a vertex $v$ on plane $P'$,
we form the contractible hull of $v$ itself; this consists of all planes
passing through $v$ and in particular $P'$.  When we do this, we remove
from $TL$ and $TR$ any triangle using $v$ as a vertex.
Once all remaining vertices are disjoint from $P'$, all the triangles are
disjoint from each other.  We then arbitrarily choose pairs of
triangles, one from $TL$ and one from $TL$, until we run out of
triangles in one of the two sets.  Each of the pairs gives a six-site
set with contractible hull containing $P'$, because the triangle above
$P'$ and the triangle below $P'$ project to sets with intersecting convex
hulls: specifically, their intersection contains the point~$c'$.
\end{proof}

\section{NP-hardness}

We now briefly discuss the computational complexity of testing the
regression depth or contractible partition number for a given plane.
Clearly, when the dimension is a fixed constant, the regression depth
can be tested in time $O(n^{d-1}+n\log n)$: there are $O(n^{d-1})$
combinatorially distinct vertical hyperplanes, the set of these vertical
hyperplanes can be constructed by forming a arrangement in a space dual
to the
$(d-1)$-dimensional projection of the points, and the number of points in
each double wedge defined by a vertical hyperplane and the input
hyperplane can be found in constant time by walking from cell to cell in
this dual arrangement.  Standard $\epsilon$-cutting methods
\cite{MulSch-HDCG-97} can be  used to design an algorithm to approximate
the regression depth  within a $(1+\epsilon)$ factor, in linear time for
any fixed values  of $\epsilon$ and the dimension.

When the dimension is
not a fixed contant, testing whether the location depth of a point is at
least some fixed bound is coNP-complete \cite{JohPre-TCS-78}.
Teng \cite{Ten-PhD-91} showed that the special case of testing whether a
point is a center point is still coNP-complete.

\begin{theorem}
Testing whether a hyperplane has regression depth at least $n/(d+1)$ is
coNP-complete.
\end{theorem}

\begin{proof}
First, to show that a hyperplane does not have high regression depth,
we need merely exhibit a double wedge bounded by it and a vertical
hyperplane that contains few points.  Therefore, the problem of testing
regression depth is in coNP.

If one could compute regression depth, one could use this to
compute the location depth of a point $x$ by finding a projective
transformation taking $x$ to $\hat\infty$ and testing the regression
depth of the image of the hyperplane at infinity.  This transformation
is a reduction from testing center points to testing regression depth;
therefore testing regression depth is coNP-complete.
\end{proof}

Therefore, also, computing the regression depth of a hyperplane is
NP-hard, since one could test regression depth by comparing the computed
depth to the value $n/(d+1)$.  However, these results do not rule out the
possibility of an efficient algorithm for finding a deep hyperplane.

Teng \cite{Ten-PhD-91} also showed that the problem of testing whether
the Tverberg depth of a point is at least some fixed bound, or of
testing whether the point is a Tverberg point, is NP-complete.
Using the same transformational ideas as before, this leads
immediately to the following result:

\begin{theorem}
Testing whether a hyperplane has contractible partition number at
least $n/(d+1)$ is NP-complete.
\end{theorem}

The computational complexity of computing a deep hyperplane or a
hyperplane with high contractible partition number remains open.

\subsection*{Acknowledgements}

Work of Amenta, Eppstein, and Teng was performed in part while visiting
the Xerox Palo Alto Research Center.
David Eppstein's work was supported in
part by NSF grant CCR-9258355 and by matching funds from Xerox Corp.
Shang-Hua Teng's work was supported in part by an Alfred P. Sloan 
Fellowship.

The authors would like to acknowledge helpful conversations with
Zbigniew Fiedorow, Harald Hanche-Olsen, Dan Hirschberg, Mia Hubert,
Peter Rousseeuw, Jack Snoeyink, Rafe Wenger, and Fran\-ces Yao.

\bibliography{rousseeuw}
\end{document}